# Germanene termination of Ge$_2$Pt crystals on Ge(110)

P. Bampoulis, L. Zhang, A. Safaei, R. van Gastel, B. Poelsema and H.J.W. Zandvliet

Physics of Interfaces and Nanomaterials group, MESA+ Institute for Nanotechnology and University of Twente, P.O. Box 217, 7500AE Enschede, The Netherlands

**Abstract**

We have investigated the growth of Pt on Ge(110) using scanning tunneling microscopy and spectroscopy. The deposition of several monolayers of Pt on Ge(110) followed by annealing at 1100 K results in the formation of three-dimensional metallic Pt-Ge nanocrystals. The outermost layer of these crystals exhibits a honeycomb structure. The honeycomb structure is composed of two hexagonal sub-lattices that are displaced vertically by 0.2 Å with respect to each other. The nearest-neighbor distance of the atoms in the honeycomb lattice is 2.5±0.1 Å, i.e. very close to the predicted nearest-neighbor distance in germanene (2.4 Å). Scanning tunneling spectroscopy reveals that the atomic layer underneath the honeycomb layer is more metallic than the honeycomb layer itself. These observations are in line with a model recently proposed for metal di-(silicides/)germanides: a hexagonal crystal with metal layers separated by semiconductor layers with a honeycomb lattice. Based on our observations we propose that the outermost layer of the Ge$_2$Pt nanocrystal is a germanene layer.

In 2004 Novoselov and Geim [1] ignited a revolution in materials science by preparing graphene, i.e. a single layer of *sp$^2$* hybrizided carbon atoms. The unique electronic structure of this archetype 2D material has led to a large number of exciting physical discoveries [2-4]. Shortly after this discovery it has been suggested that two-dimensional sheets of other group IV elements, such as Si [5,6] and Ge [6], might exhibit similar properties as graphene. Already in 1994 Takeda and Shiraishi [7] performed quantum mechanical *ab initio* calculations on planar silicon and germanium structures that have the graphite structure. They pointed out that the lowest energy configuration was obtained if the two atoms of the honeycomb are slightly displaced with respect to each other in a direction normal to the planar structure. Their calculations also revealed that these buckled Si and Ge structures exhibited semi-metallic properties. Unfortunately, they did not pay any attention to the exact *k*-dependence of the energy dispersion relations in the vicinity of the Fermi level. More than a decade later Guzmán-Verri and Lew Yan Voon [5] showed, using tight binding calculations, that a 2D silicon sheet with the graphite structure has Dirac cones. Hence the electrons in these 2D silicon sheets behave as massless Dirac fermions. The Si and Ge analogues of graphene are referred to as silicene and germanene, respectively. First–principles calculations by Cahangirov et al. [6] revealed that a single sheet of germanium atoms with a honeycomb structure is also stable. The free-standing Ge honeycomb lattice is not fully planar, but buckled. The two hexagonal sub-lattices of the honeycomb lattice are displaced vertically by 0.64 Å, which is slightly larger than the buckling in silicene (0.44 Å). The buckling results into a weaker $\pi$ bonding and the perpendicular $p_z$ orbital hybridizes with the in-plane orbitals. Similar to graphene and silicene the $\pi$-$\pi^*$ band crossings at the $K$ and $K'$ points of the hexagonal Brillouin zone occur at the Fermi level. In the vicinity of the Fermi level the $E(k)$ dispersion relations are linear and therefore the charge carriers behave as massless Dirac fermions. The calculated Fermi velocity of germanene is, however, smaller than the Fermi velocity of graphene. Due to germanium's large atomic number the spin-orbit coupling is rather large. The spin-orbit gap of the $\pi$ orbitals of germanene at the Dirac points is 23.9 meV [8], which is substantially larger than the spin-orbit gaps of graphene (< 0.1 meV) and silicene (1.55 meV). The large spin-orbit gap makes germanene an excellent candidate to exhibit the quantum spin Hall effect [8]. The existence of this quantum spin Hall effect was put forward by Kane and Mele [9]. The quantum spin Hall effect state is characterized by a bulk gap and conducting spin-polarized edge states without dissipation at the sample boundaries. The quantum spin Hall effect in graphene is very difficult to access because of the required extremely low temperatures. In the case of germanene, however, the quantum spin Hall effect should occur near room temperature.

In contrast to graphene, silicene and germanene do not occur in nature and therefore these materials have to be synthesized. Silicene and silicene nanoribbons have already been grown on Ag(110), Ag(111), ZrB$_2$(0001) and Ir(111) surfaces [10-16], but to the best of our knowledge there are only two papers that report the growth of germanene [17,18]. Dávila et al. [17] have studied the growth germanene on Au (111) using scanning tunneling microscopy, synchrotron radiation core-level spectroscopy and density functional theory calculations. Li et al. [18] have grown a single layer of germanium on a Pt(111) substrate and subsequently analyzed this germanium sheet with low energy electron diffraction and scanning tunneling microscopy. The low energy electron diffraction pattern reveals a ($\sqrt{19}$x$\sqrt{19}$) reconstruction that is slightly rotated with respect to the underlying Pt(111) lattice. These observations are confirmed by scanning tunneling microscopy images, albeit Li et at. [18] did not manage to achieve atomic resolution. However, in a recent paper by Švec et al. [19] it was suggested that the Ge induced ($\sqrt{19}$x$\sqrt{19}$) reconstruction on Pt(111) is in fact a surface alloy composed of Ge$_3$Pt tetramers that resembles a twisted kagome lattice.

Here we study the growth of Pt on a Ge(110) substrate. The deposition of a few monolayers of Pt on Ge(110) followed by annealing at temperatures around 1100 K leads to the formation three-dimensional nanocrystals on the Ge(110) substrate. These three-dimensional crystals are terminated by a buckled honeycomb structure, which we interpret as germanene.

Experiments have been performed with a scanning tunneling microscope operating in ultra-high vacuum. The Ge(110) samples are cut from nominally flat 10x10x0.5 mm, single-side-polished $n$-type substrates. Samples are mounted on Mo holders and contact of the samples to any other metal during preparation and experiment has been carefully avoided. The Ge(110) samples have been cleaned by 800 eV Ar$^+$ ion sputtering and subsequent annealing at 1100 (±25) K, a procedure that is commonly used to prepare atomically clean Ge(001) and Ge(111) surfaces [20]. Subsequently, several monolayers of Pt are deposited onto the surface at room temperature. Pt is evaporated by resistively heating a W wire wrapped with high purity Pt (99.995%). After Pt-deposition the sample is flash annealed at 1100 (±25) K and subsequently cooled down to room temperature before placing it into the scanning tunneling microscope for observation.

In Fig. 1(A) a large-scale scanning tunneling microscopy image of the bare Ge(110) surface is shown. In contrast to its low index cousins the unreconstructed Ge(110) is anisotropic. The bulk terminated Ge(110) surface consists of atomic trenches that run in the [1-10] direction. The clean and reconstructed Ge(110) surface is well-documented and exhibits (16x2) and c(8x10) reconstructed domains as well as disordered domains [21-24]. The elementary building block of all these reconstructions is a pentagon, a five-membered Ge atom ring. These pentagons show up in the inset of Fig. 1(A) as small bright dots, where a (16x2) reconstructed domain is shown. In Fig. 1(B) a scanning tunneling microscopy image taken after the deposition of 2-3 monolayers of Pt and annealing at 1100 K is depicted. The surface is composed of flat regions (ii) and three-dimensional crystals (i). In Fig. 1(C) a scanning tunneling microscopy image recorded on-top of one of the three-dimensional crystals is depicted. The step edges are very straight and their heights are quantized in units of 5.6±0.6 Å (see Fig. 1(D)). Scanning tunneling spectra recorded on both regions reveal that the (ii) regions are semiconducting, whereas the (i) regions are metallic (see Fig. 1(E)).

It is important to point out that the formation of these three-dimensional crystals only occurs upon annealing at temperatures higher than 1000-1100 K. The phase diagram of Pt-Ge system exhibits an eutectic at 1043 K. This occurs at a composition of 22 % and 78 % Pt and Ge respectively. Low energy electron microscopy (LEEM) images show that slightly above this temperature liquid drops are formed and move as large entities across the surface [25]. A LEEM movie of the formation and diffusion of these eutectic droplets is included as supplementary material. The eutectic droplets cover approximately 5 % of the surface. Upon cooling down these droplets solidify and spinodal decomposition must be expected to occur. Ideally, two stable coexisting phases will form, which are most close to the eutectic composition. The (bulk-) phase diagram identifies a low density or even platinum-free phase and an ordered $Ge_2Pt$ phase. We anticipate that the cluster (i) in Fig. 1 (B) is actually a $Ge_2Pt$ crystallite. To obtain more information we decided to have a more detailed look at the structure of the crystallites' outermost layer. In Fig. 2 a series of scanning tunneling microscopy images recorded at different voltages on the three-dimensional crystals are shown. A well-ordered honeycomb lattice, which consists of two hexagonal sub-lattices is resolved. At +0.7 V only one of these two hexagonal sub-lattices is visible. In Fig. 3 a filled state scanning tunneling microscopy image (3D representation) is shown. Scanning tunneling spectra are recorded on the atoms of the honeycomb, as well as in the center of the honeycomb (see Fig. 3(B)). Both spectra are recorded at the same set point and it is evident that the center of the honeycomb is more metallic than the atoms of the honeycomb. The height difference between the brightest and dimmest atom of the honeycomb is 0.2 Å, the nearest-neighbor distance is 2.5±0.1 Å and the honeycomb lattice constant is 4.4±0.2 Å. Aside from the fact that these lattice parameters are in favor of germanene we would like to emphasize that Pt tends to maximize its total number of neighbors and therefore the honeycomb lattice cannot consist of Pt atoms alone.

Our findings are intimately in line with a structural model published by Xie and Nesper [26] for $SrGe_{1.2}Si_{0.8}$. They found a hexagonal (P6/mmm) structure with lattice vectors *a*=0.436 and *c*=0.454 nm. Very similar values were obtained for $CaSi_{1.2}Zn_{0.8}$. A model of this structure is shown in Fig. 3 (C): the metal atoms (Sr,Ca) are located in one plane in between planes with the semiconductor atoms with a honeycomb structure. The atomic density of the layer of semiconductor atoms (two atoms per honeycomb unit cell) is twice the atomic density of the Pt layer (one atom per honeycomb unit cell, see Fig. 3C). The Ge layer is the terminating plane with its lower surface free energy and the natural tendency of the metal (Pt) atoms to search for the highest coordination. The IV curves measured in the center of the honeycomb mainly probe the second (Pt) layer, while the IV curves taken at the atoms of the honeycomb probe the top germanene layer (see Fig. 3(B)). Further support for this attribution is coming from the *a* parameter of the lattice taken from Fig. 2 (B), equaling 0.44 nm, in perfect agreement with Ref. [26]. From the line profile in Fig. 1 (D) we extract a step height of 0.56±.06 nm. Note that we also observe a double step in the line profile. Regardless of the termination this determines a *c/a* ratio of 1.29, i.e. slightly larger than the value of 1.05 for the cases of Ref. [26]. Since the ionic radius of Pt is smaller than both that of Ca and of Sr we conclude that the bond strength of the semiconductor plane and the metallic planes is less strong. Therefore, the terminating semiconductor planes incline towards an enhanced germanene character. This is further reinforced by the fact that, as shown in Figs. 2 and 3, some atoms of the honeycomb appear a bit brighter than others. At small negative sample biases and positive sample biases it is clear that the honeycomb lattice consist of two hexagonal sub-lattices that are slightly displaced vertically with respect to each other. This is exactly what one would expect for germanene, albeit the observed buckling of 0.2 Å is still substantially smaller than the predicted buckling for free-standing germanene (0.64 Å). Therefore, we think we have a strong case for a germanene termination of the (0001) surface of $Ge_2Pt$ crystallites on Ge(110).

In conclusion, the deposition and subsequent annealing of Pt on Ge(110) leads to the formation of microscopic Pt/Ge crystals. The outermost layer of these crystals exhibits a honeycomb structure. This honeycomb lattice is composed of two hexagonal sub-lattices that that are displaced vertically by 0.2 Å with respect to each other. The nearest-neighbor distance is 2.5±0.1 Å. Based on our experimental observations we propose that the outermost layer of the Pt/Ge crystals is a germanene layer.

**Acknowledgements**

P. Bampoulis and A. Safaei would like to thank the Dutch Organization for Research (NWO, STW 11431) and the Stichting voor Fundamenteel Onderzoek der Materie (FOM, 10ODE04) for financial support. L. Zhang acknowledges the China Scholarship Council (CSC) for financial support.

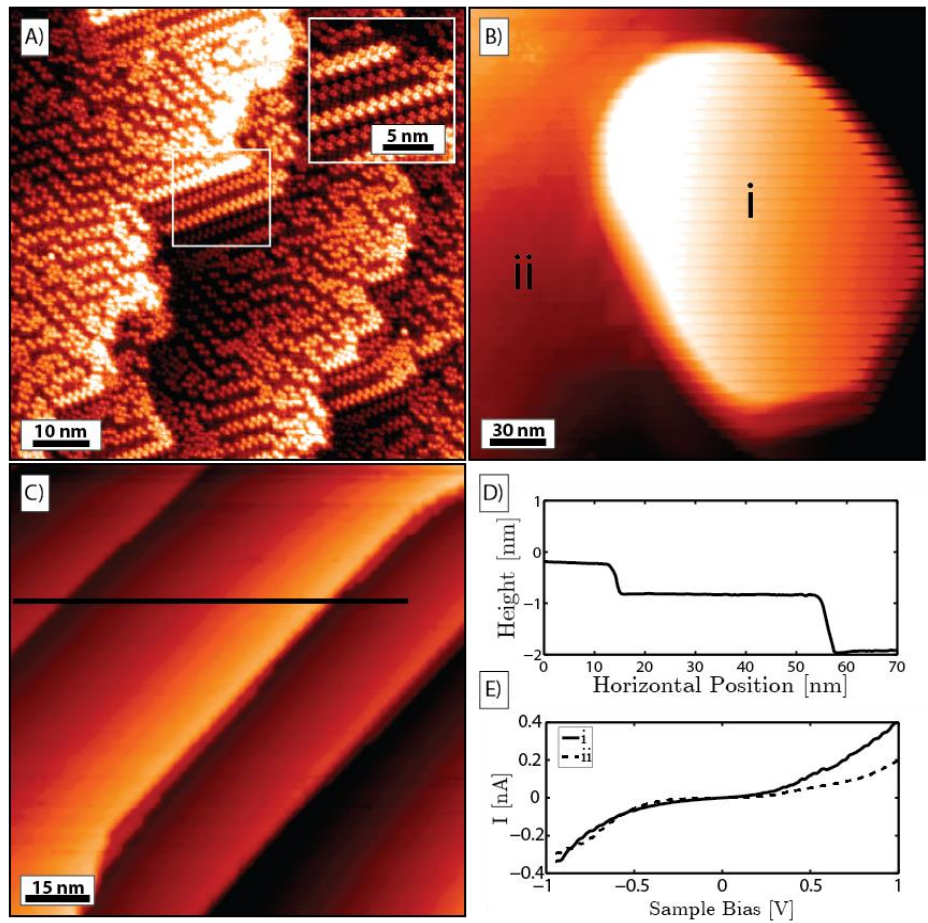

**Figure 1**
(A) Scanning tunneling microscopy image of a Ge(110) surface. Sample bias -1.0 V and tunneling current 1.0 nA. Inset: (16x2) reconstruction. Sample bias -1.0 V and tunneling current 0.5 nA. (B) Scanning tunneling microscopy image after deposition and annealing at 1100 K of Pt on Ge(110). Sample bias -1.0 V and tunneling current 0.5 nA (please note that the regularly spaced lines on the cluster are due to the grid IV scan). The height of the cluster is ∼20 nm and the aspect ratio 3:4. (C) Scanning tunneling microscopy image recorded on the cluster (region i) shown in (B). Sample bias -1.4 V and tunneling current 0.5 nA. (D) Line scan taken across the step edges in (C). (E) IV curves recorded on regions (i) and (ii). Set points: Sample bias -1.0 V and tunneling current 0.5 nA.

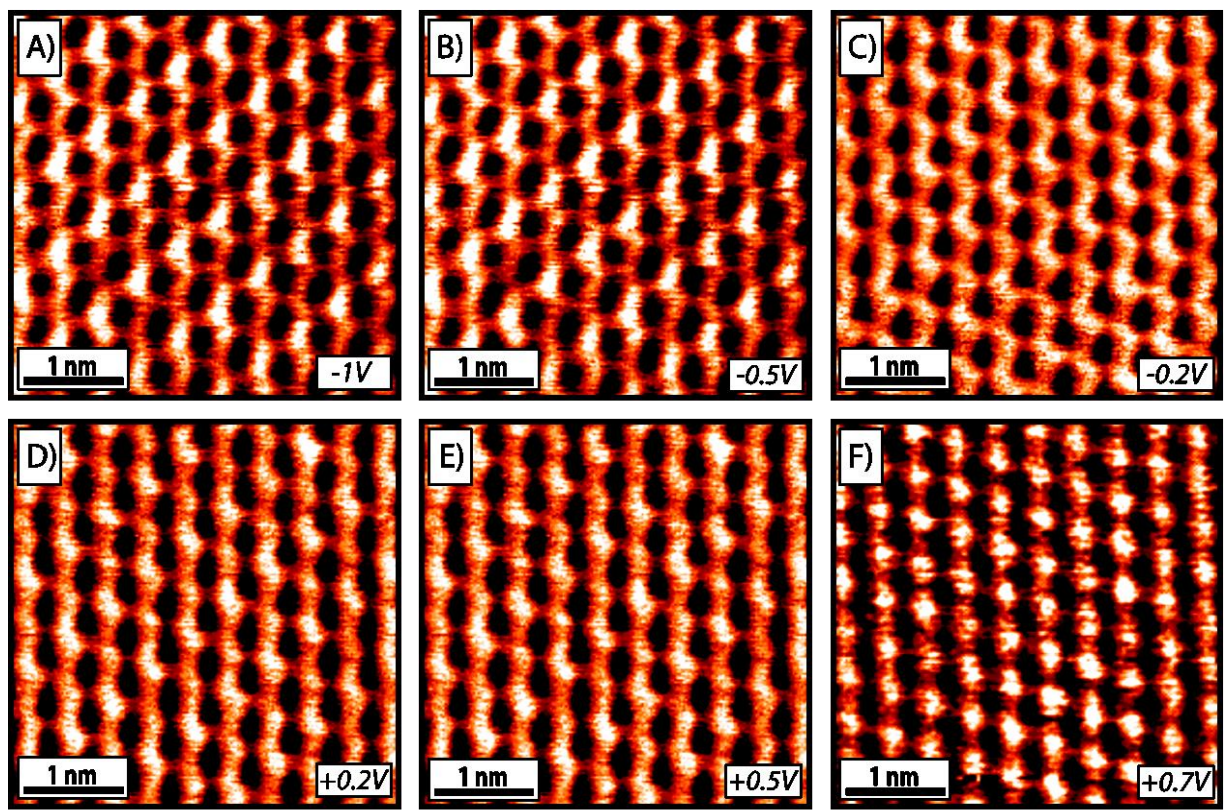

**Figure 2**
Scanning tunneling microscopy images taken at -1 V (A), -0.5 V (B), -0.2 V (C), +0.2 V (D), +0.5 V (E) and +0.7 V (F). Tunnel current is 0.5 nA (A) and 0.2 nA (B-F). Image size 3.9 nm x 3.9 nm.

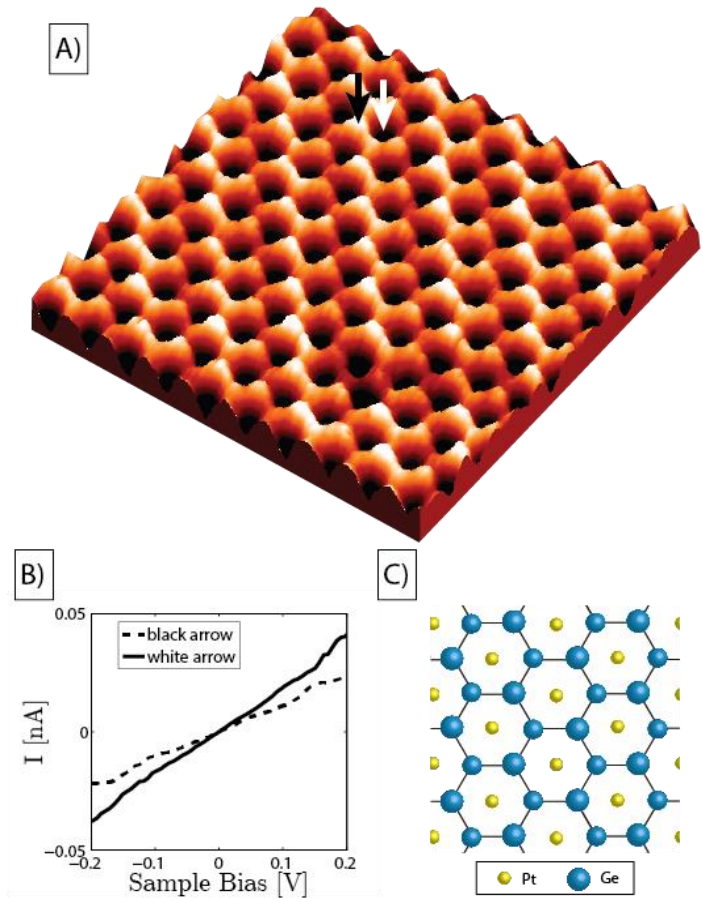

**Figure 3**

(A) Scanning tunneling microscopy image of the honeycomb lattice. Sample bias -0.5 V, tunnel current 0.2 nA. Image size 4.5 nm x 4.5 nm. (B) I(V) curve recorded at a bright atom of the honeycomb lattice (dotted line, black arrow) and at the center of the honeycomb (solid line, white arrow). (C) Schematic ball model. Top layer: buckled honeycomb lattice (germanene, in blue) second layer: platinum (in yellow).